\documentclass[11pt,english]{article}
\usepackage{graphicx}
\usepackage{amstext}
\usepackage{caption}
\usepackage{etoolbox}
\usepackage{makeidx}
\include{epsf}
\usepackage{amsfonts}
\usepackage{authblk} 
\usepackage{sectsty}
\usepackage{amsmath,amssymb,epsfig}
\usepackage{amscd}
\usepackage{amsthm}
\usepackage{mathrsfs}
\usepackage{dsfont}
\usepackage[applemac]{inputenc}
\usepackage[english]{babel}
\usepackage{enumitem} 
\usepackage[]{latexsym}
\usepackage{caption}
\usepackage{subfigure}
\usepackage{hyperref}
\usepackage{blindtext}
\usepackage{verbatim}
\usepackage{cancel}

\hypersetup{
pdftitle={},%
pdfauthor={},%
pdfsubject={},%
pdfkeywords={},%
colorlinks=true,%
linkcolor=blue,%
citecolor=red,%
linktocpage=true,%
hyperfootnotes=true,%
pageanchor=true
}

\newcommand*{\affaddr}[1]{#1}
\newcommand*{\affmark}[1][*]{\textsuperscript{#1}}

\newtheorem*{proof*}{Proof}

\newcommand{\be}{\begin{equation}}

\newcommand{\ee}{\end{equation}}
\def\beqa{\begin{eqnarray}}
\def\eeqa{\end{eqnarray}}
\def\bean{\begin{eqnarray*}}
\def\eean{\end{eqnarray*}}

\renewenvironment{thebibliography}[1]
         {\section*{References}\frenchspacing\small
          \begin{list}{[\arabic{enumi}]}
         {\usecounter{enumi}\parsep=2pt\topsep 0pt
         \settowidth{\labelwidth}{[#1]}
         \leftmargin=\labelwidth\advance\leftmargin\labelsep
         \rightmargin=0pt\itemsep=1pt\sloppy}}{\end{list}}

 \numberwithin{equation}{section}

\input xy
\xyoption{all}

\usepackage[normalem]{ulem}

\newcommand{\braket}[2]{\left\langle \vphantom {#1 #2} #1 \hphantom{|} \right| \left. \vphantom {#1 #2} #2 \right\rangle}

\title{\textbf{\textsf{Is limiting curvature mimetic gravity an effective polymer quantum gravity?}}\vspace{0.35cm}}

\author{
\textsf{Norbert Bodendorfer\affmark[1]\footnote{\texttt{norbert.bodendorfer@physik.uni-r.de}}, Fabio M. Mele\affmark[1]\footnote{\texttt{fabio.mele@physik.uni-r.de}}, and Johannes M\"unch\affmark[1]\footnote{\texttt{johannes.muench@physik.uni-r.de}}}\\
\affaddr{\affmark[1]\textsf{Institute for Theoretical Physics, University of Regensburg,}}\\
\affaddr{\textsf{93040 Regensburg, Germany}}\vspace{-0.5cm}
}

\begin{document}

\maketitle

\begin{abstract}
\textsf{A recently proposed version of mimetic gravity incorporates a limiting curvature into general relativity by means of a specific potential depending on the d'Alembertian of the scalar field. In the homogeneous and isotropic setting, the resulting theory agrees with the simplest incarnation of effective loop quantum cosmology (LQC) once the limiting curvature is identified with a multiple of the Planck scale. In this paper, we answer the question of whether such a relation can hold in the context of Bianchi I models. Our result is negative: it turns out to be impossible to view the Hamiltonian of limiting curvature mimetic gravity as an effective LQC Hamiltonian due to the appearance of terms that cannot be supported on the polymer Hilbert space underlying LQC. The present analysis complements a related result in the context of spherical symmetry. }
\end{abstract}

\section{Introduction}

Mimetic gravity \cite{ChamseddineMimeticDarkMatter} has been proposed as a modified version of general relativity incorporating cold dark matter, see \cite{SebastianiMimeticGravityA} for an overview. A specifically interesting variant \cite{ChamseddineResolvingCosmologicalSingularities} incorporates a limiting curvature by means of a specific potential depending on the d'Alembertian of a scalar field, the only field entering the action beyond the metric. While the form of the action was argued to be motivated by non-commutative geometry in \cite{ChamseddineResolvingCosmologicalSingularities}, the potential was chosen by hand. However, by relating the resulting theory to homogeneous and isotropic loop quantum cosmology, see \cite{BodendorferCanonicalStructureOf, NouiEffectiveLoopQuantum}, one can make a simplicity argument \cite{BodendorferCanonicalStructureOf} to obtain the potential from quantum mechanics.

Given this situation, it is interesting to ask whether such a relation can hold more generally, i.e. by dropping the symmetry assumptions of homogeneity or isotropy. In particular, it was stressed in \cite{BodendorferCanonicalStructureOf, NouiEffectiveLoopQuantum} that one could then view limiting curvature mimetic gravity or generalised versions of it as effective actions for loop quantum gravity. This is on the one hand particularly interesting for applications of loop quantum gravity to more complicated systems such as black holes, where working directly within the full theory might be too involved computationally. On the other hand, it may shed light on the ongoing debate on possible modifications of general covariance in loop quantum gravity, see e.g. \cite{BojowaldEffectiveLineElements}, as mimetic gravity is manifestly covariant.

The most convenient setting to check the desired relation are Bianchi I cosmological models which are homogeneous, but non-isotropic. Here, the cosmological solutions in \cite{ChamseddineResolvingCosmologicalSingularities} match those of loop quantum cosmology \cite{GuptQuantumGravitationalKasner} qualitatively and thus suggest that equivalence between the two theories may hold for a specific (polymerisation) choice in the loop quantum cosmology dynamics. It turns out that we are able to establish a no-go result based on an analytic computation as follows: we first bring limiting curvature mimetic gravity in a form suitable to be compared with loop quantum cosmology via a variable transformation. By expressing the mimetic gravity Hamiltonian in terms of those variables, we can check whether the resulting expression can be supported as an operator on the Hilbert space underlying loop quantum cosmology. This turns out to be impossible for the potential chosen in \cite{ChamseddineResolvingCosmologicalSingularities}, but also for general potentials by a simple argument. 
Our analysis complements a related computation that has been performed in the context of spherical symmetry \cite{AchourNonSingularBlack}.

This paper is organised as follows:\\
Section \ref{sec:Review} reviews previous work and introduces the relevant concepts. The main computation is done in section \ref{sec:Computation}. A brief conclusion is given in section \ref{sec:Conclusion} and a computation omitted in the main part is detailed in the appendix.

\section{Review} \label{sec:Review}

\subsection{Conventions}

In what follows, we will use the Einstein sum convention for repeated indices when covariant and contravariant indices are contracted, while no sum is meant if the repeated indices are all up or down. The signature of the metric is $(-, +, +, +)$. We set $8 \pi G = 1$. For simplicity, we choose an orientation of the spatial slice so that all scale factors are positive. We will restrict here to a 3-torus topology of the spatial slices with unit coordinate volume for simplicity, but all our arguments apply more generally by taking into account fiducial cells. We denote spatial tensor indices by $a,b$ when no symmetry assumptions are present and by $i,j$ in the context of Bianchi I.

\subsection{Limiting curvature mimetic gravity: Canonical structure}

The class of models leading to limiting curvature mimetic gravity is defined by the action \cite{ChamseddineResolvingCosmologicalSingularities}
\be
	S = \int d^4x \, \sqrt{-g} \left( \frac{1}{2} R + \frac{1}{2} \lambda \left(1+ g^{\mu \nu} \partial_\mu \phi \partial_\nu \phi \right) + f (\Box \phi)\right) \label{eq:MCAction} \text{,}
\ee
where we note that $\lambda$ enforces the constraint $g^{\mu \nu} \partial_\mu \phi \partial_\nu \phi = -1$. The potential $f (\Box \phi)$ is a priori arbitrary. A detailed canonical analysis of this action was performed in \cite{BodendorferCanonicalStructureOf}, see also \cite{KlusonCanonicalAnalysisOf} for previous work. It is convenient to first bring the action into the more suitable form 
\begin{align}
	S &= \int d^4x \, \sqrt{-g} \left( \frac{1}{2} R + \frac{1}{2} \lambda \left(1+ g^{\mu \nu} \partial_\mu \phi \partial_\nu \phi \right) + f (\chi) + \beta (\chi - \Box \phi)\right) 
\end{align}
by introducing the auxiliary fields $\beta$ and $\chi$. The resulting canonical theory is rather cumbersome and lengthy to state explicitly, which is why we will only display the most relevant parts here.

The Hamiltonian is, as usual, a sum of constraints, most notably the Hamiltonian constraint
\begin{align}\label{CMH}
\mathcal{H} =&\; \frac{2}{\sqrt{q}} P^{a b} P^{c d} \left( q_{ac} q_{bd} - \frac{1}{2} q_{a b} q_{cd} \right) - \frac{1}{2} \sqrt{q} R^{(3)} - \frac{1}{2} \lambda \sqrt{q} \left( 1 + q^{a b} \partial_a \phi \partial_b \phi - \frac{P^2_\beta}{q} \right)
\notag
\\
&-\sqrt{q} \left(f(\chi) + \beta \chi\right)-\sqrt{q} q^{a b} \partial_a \beta \partial_b \phi - \frac{1}{\sqrt{q}} P_\phi P_\beta \text{,}
\end{align}
where $P_\beta$ and $P_\chi$ are the conjugate momenta of $\beta$ and $\chi$. Due to the appearance of second class constraints related to the auxiliary fields (see below), the resulting Dirac brackets are quite involved. For this paper, it will be sufficient to only consider 
\begin{align}
&\;\left\{q_{ab}(x),q_{cd}(y)\right\}_*=0\;,\label{DB2}\\
&\left\{q_{ab}(x),P^{cd}(y)\right\}_*\overset{\partial_a\phi=0}{=}\delta^{(3)}(x,y)\left(\delta_{(a}^c\delta_{b)}^d-q_{ab}q^{cd}\,\frac{f''(\chi)}{3f''(\chi)-2}\right)\;,\label{DB}\\
&\left\{P^{ab}(x),P^{cd}(y)\right\}_*\overset{\partial_a\phi=0}{=}-\delta^{(3)}(x,y)\left(q^{ab}P^{cd}-q^{cd}P^{ab}\right)\frac{f''(\chi)}{3f''(\chi)-2}\;\label{DB1}\text{.}
\end{align}
where we use the notation $\{\cdot,\cdot\}_*$ for the Dirac bracket and by $\overset{\partial_a\phi=0}{=}$ we mean equality up to terms containing spatial gradients of $\phi$. This restriction was chosen in \cite{BodendorferCanonicalStructureOf} for convenience, as otherwise the Dirac bracket becomes too intricate. In the Bianchi I models considered in this paper, it is automatically satisfied.

An interesting choice for $f(\chi)$ is given by 
\be\label{fCM}
f(\chi)=\chi_m^2\left(1+\frac{1}{3}\frac{\chi^2}{\chi_m^2}-\sqrt{\frac{2}{3}}\frac{\chi}{\chi_m}\arcsin\left(\sqrt{\frac{2}{3}}\frac{\chi}{\chi_m}\right)-\sqrt{1-\frac{2}{3}\frac{\chi^2}{\chi_m^2}}\,\right)\;.
\ee
which results in singularity-free cosmological models \cite{ChamseddineResolvingCosmologicalSingularities} and Schwarzschild black holes \cite{ChamseddineNonsingularBlackHole} with limiting curvature $\sim \chi_m^2$. This choice can also be extracted from the simplest choice of Hamiltonian in loop quantum cosmology by an inverse Legendre transform \cite{DateEffectiveActionsFrom}. We will adopt this choice in the paper for concreteness, but our argument will in the end generalise to arbitrary $f$.

\subsection{Limiting curvature mimetic gravity: Bianchi I sector}

Bianchi I models are the simplest homogeneous but non-isotropic cosmological models. In a suitable coordinate system $(t,x^i)$, the metric reads 
\be\label{3metric}
q_{i j} = q_{i i} \delta_{i j}, \quad q_{i i} = a_{i}^2\;,
\ee
where the $a_i$ are the three scale factors. Similarly, we have
\be\label{PCM}
P^{i j} = P^{i i} \delta^{i j}\;.
\ee
This metric is spatially flat, i.e. $R^{(3)} = 0$. Scalar fields consistent with this symmetry only have a time dependence, i.e. $\partial_a\phi = 0$. For unit lapse, the above constraint $g^{\mu \nu} \partial_\mu \phi \partial_\nu \phi = -1$ leads to $\phi = \pm t + const$. 

It is convenient to solve the second class constraints appearing in the above canonical analysis for $\beta, P_\beta$, and $\chi$ in our symmetry reduced context. Without stating their precise general form (see \cite{BodendorferCanonicalStructureOf}), we obtain 
\begin{align}
&C_\lambda = 0\qquad \Rightarrow \qquad P^2_\beta = q\;,
\\
&C_\chi = 0 \qquad \Rightarrow \qquad \beta = - f'(\chi)\;,
\\
&D_\lambda = 0 \qquad \Rightarrow \qquad \chi = - \frac{P}{\sqrt{q}} = - \frac{1}{\sqrt{q}}q_{ii} P^{ii}\;.
\label{chi}
\end{align}
Note that in the Bianchi I case $P= \frac{\sqrt{q}}{2}q_{ab} (K^{ab} - q^{ab} K)=-\stackrel{\textbf{\;\;\;\,.}}{\sqrt{q}}$, where $K_{ab}$ is the extrinsic curvature and the dot denotes derivatives w.r.t. the time coordinate (with unit lapse) to avoid confusion with primes, which instead denote generic derivatives w.r.t. the argument of the function. Hence, Eq. \eqref{chi} relates $\chi$ to the expansion of spacetime, namely $\chi=\stackrel{\textbf{\;\;\;\,.}}{\sqrt{q}}/\sqrt{q}$. Inserting all this in the Hamiltonian \eqref{CMH} and performing the trivial spatial integrations gives
\begin{align}
\mathcal{H} = \frac{2}{\sqrt{q}} P^{a b} P^{c d} \left( q_{ac} q_{bd} - \frac{1}{2} q_{a b} q_{cd} \right) -\sqrt{q} \left(f(\chi) - f'(\chi) \chi\right) \pm P_\phi\;.
\end{align}
To compare with loop quantum cosmology, we focus on the gravitational part (dropping $P_\phi$)
\begin{align}
\mathcal{H}_{\text{grav}} =&\; \sqrt{q} \left(\chi^2 -f(\chi) + f'(\chi) \chi \right) -\frac{4}{\sqrt{q}}\left(q_{11} P^{11} q_{22} P^{22} + q_{11} P^{11} q_{33} P^{33}+ q_{22} P^{22} q_{33} P^{33}\right)\;.
\label{HCM3}
\end{align}

\subsection{Loop quantum cosmology}

There are several ways to arrive at loop quantum cosmology, either by quantising a symmetry reduced theory as, e.g., in \cite{AshtekarMathematicalStructureOf, AshtekarQuantumNatureOf} or by considering reduction of full loop quantum gravity \cite{BIII, BVI, OritiBouncingCosmologiesFrom, AlesciImprovedRegularizationFrom}. Here, we will take the first route and present a simplified derivation by direct polymerisation of the symmetry reduced classical Hamiltonian. 

Following Arnowitt, Deser, and Misner \cite{ArnowittTheDynamicsOf}, classical canonical general relativity is usually written using the spatial metric $q_{ab}$ and its conjugate $P^{ab} = \frac{\sqrt{q}}{2} (K^{ab} - q^{ab} K)$, where $K_{ab}$ is the extrinsic curvature. In the context of a Bianchi I model as above, $q_{i j} = \text{diag}(a_1^2, a_2^2, a_3^2)_{i j}$ and one may consider the variables 
\be
	p_1= a_2a_3, ~~ p_2= a_1a_3, ~~ p_3= a_1a_2, \label{eq:ps} 
\ee
as well as
\be
	c^1 = K_{11} / a_1, ~~ c^2 = K_{22} / a_2, ~~ c^3 = K_{33} / a_3
\ee 
with the Poisson brackets $\{c^i, p_j\} = \delta_i^j$. The Hamiltonian constraint can be written as 
\be
\mathcal{H}_{\text{grav}} =-\sqrt{p_1 p_2 p_3} \left(\frac{ c^2 c^3}{p_1} +\frac{ c^3 c^1}{p_2} +\frac{ c^1 c^2}{p_3}   \right) \text{.} \label{eq:BianchiIH}
\ee
Another way to arrive at this formulation is to start with the SU$(2)$ Ashtekar-Barbero variables of $3+1$-dimensional canonical general relativity \cite{AshtekarNewVariablesFor, BarberoRealAshtekarVariables}, an SU$(2)$ connection $A^i_a$ and its conjugate, the densitised triad $E_i^a$. $c^i$ and $p_j$ are related to them as
\be\label{LQCvar}
A_a^i=c^i(L_i)^{-1}{\,}^{o}{\omega}_a^i,\qquad E_i^a=p_iL_iV_o^{-1}\sqrt{{}^oq}{\,}^oe_i^a\;,
\ee
where $L_i$, $i=1,2,3$, denote the lengths of the edges of the elementary cell in the three spatial directions, $V_o=L_1L_2L_3$ its volume measured by using the fiducial flat metric ${}^oq_{ab}$, and ${}^oe_i^a$, ${}^o\omega^i_a$ the fiducial triads and co-triads dual to them.

The mathematical structure of the Hilbert space underlying loop quantum cosmology is inspired by loop quantum gravity \cite{RovelliQuantumGravity, ThiemannModernCanonicalQuantum}, which in turn can be described as a diffeomorphism invariant extension of lattice gauge theory, where the (dynamical) lattice encodes gravity. As in lattice gauge theory, the scalar product of the theory is obtained by integrating an exponentiated connection against the Haar measure of the underlying group. In the cosmological setting, this translates into considering point-holonomies of $c^i$. More precisely, the Hilbert space is the completion of the functions 
\be
	\psi(c) = \sum_{\rho \in \mathbb R} \sum_{j=1}^3 \alpha^j_\rho e^{i \rho c^j}, ~~~~ \alpha_{\rho}^j \in \mathbb C
\ee
w.r.t. to the norm induced by the scalar product 
\be
	\braket{\psi}{\psi'} = \sum_{\rho \in \mathbb R} \sum_{j=1}^3 \overline{\alpha^j_\rho} \, \alpha'{}^{j}_\rho \text{.}
\ee
satisfying the square integrability condition $\braket{\psi}{\psi}<\infty$, which enforces that at most countably many $\alpha_\rho^j$ can be non-zero. These functions (for a single $j$) are known as (square integrable) almost periodic functions, see e.g. \cite{SubinDifferentialAndPseudodifferential}, and the underlying compact group is the Bohr compactification of the real line, which, roughly speaking, generalises U$(1)$ to the case of non-integer representation labels $\rho$. 

The crucial point is now that while linear combinations of $e^{i \rho c^j}$ can be promoted to (multiplication) operators on this space, functions such as $c^j$ are not well defined as operators. When quantising a given system based on this Hilbert space, one thus has to approximate all phase space functions depending on $c^j$ by linear combinations of exponentials as above, a process known as polymerisation\footnote{The choice of approximant is guided by physical principles as no naive lattice-like continuum limit is available in the background-independent setting. In the case of cosmology, one chooses $\rho$ as a function of the $p_i$ so that quantum gravity effects become important near the Planck curvature while the classical theory is recovered for small curvatures \cite{AshtekarQuantumNatureOf, AshtekarLoopQuantumCosmologyBianchi}.} \cite{CorichiPolymerQuantumMechanics}. We note that this is the direct analogue to writing the lattice Hamiltonian in terms of parallel transports.   

It turns out that for large volumes, the quantum evolution of the expectation value of $\hat p_i$ and $\hat c^j$ is well approximated by solving effective classical Hamiltonian equations of motion where all operators are approximated by their classical analogues \cite{RovelliWhyAreThe}. The original classical theory is then deformed by corrections to the equations of motion in powers of $\hbar$ stemming from substituting $c^j$ by a polymerisation, e.g. $\frac{\sin(\bar \mu_j(p) c^j)}{\bar \mu_j(p)}$, for a suitable choice of $\bar \mu_j(p)$ \cite{AshtekarLoopQuantumCosmologyBianchi}. 

It is crucial to note that due to this change of Hamiltonian, the original relation of $c^j$ to time derivatives of the spatial metric is obscured, since such a relation is dictated by the Hamiltonian. Therefore, while $\{c^i, p_j\} = \delta^i_j$ still holds for the effective theory, $c^i$ is a complicated function of $\dot a_j$ depending on the chosen polymerisation scheme.

\section{Comparison with loop quantum gravity} \label{sec:Computation}

\subsection{Strategy}

Comparing the Hamiltonian formulations of loop quantum cosmology and limiting curvature mimetic gravity consists of two steps:
\begin{enumerate}
\item We look for a canonical transformation $(q_{ii},P^{ii})\mapsto(c^i(q,P),p_i(q,P))$\footnote{Here $(q,P)$ is meant just as a shorthand notation for the dependence from all the $q_{ii}$ and the $P^{ii}$ and it must not be confused with their determinants.} such that the canonical structures of the two theories agree. In other words, we demand the Dirac brackets between $c^i(q,P)$ and $p_j(q,P)$ in limiting curvature mimetic gravity to be equal to the canonical Poisson brackets of effective loop quantum cosmology
\be\label{PBDB1}
\left\{c^i(q,P),p_j(q,P)\right\}_*\overset{!}{=}\left\{c^i,p_j\right\}=\delta^i_j\;,
\ee
which, together with
\be\label{PBDB2}
\left\{c^i(q,P),c^j(q,P)\right\}_*\overset{!}{=}\left\{c^i,c^j\right\}=0\;,
\ee
provide a set of first-order partial differential equations (PDEs) in the unknowns $c^i(q,P)$, $p_j(q,P)$.
\item Once this dependence is known, we can invert these functions and insert the resulting expressions into the Hamiltonian \eqref{HCM3}. We then check whether the resulting Hamiltonian can be written as a linear combination of operators that are well defined on the corresponding loop quantum gravity Hilbert space. 
\end{enumerate}

A simplification in the problem occurs since $p_j$ as a configuration variable is unchanged by the modification in the LQC Hamiltonian and should thus only depend on $q_{ij}$ as in \eqref{eq:ps}. 

\subsection{Canonical transformation}

With the above simplification, the Dirac brackets on the l.h.s. of \eqref{PBDB1} read
\begin{align}\label{cpDB}
\left\{c^i(q,P),p_j(q)\right\}_*&=\frac{\partial c^i}{\partial q_{kl}}\{q_{kl},q_{mn}\}_*\frac{\partial p_j}{\partial q_{mn}} +\frac{\partial c^i}{\partial P^{kl}}\{P^{kl},q_{mn}\}_*\frac{\partial p_j}{\partial q_{mn}} \notag\\
&=-\frac{\partial c^i}{\partial P^{kl}}\left(\delta_{(m}^k\delta_{n)}^l-q_{mn}q^{kl}\,\frac{f''(\chi)}{3f''(\chi)-2}\right)\frac{\partial p_j}{\partial q_{mn}}\;,
\end{align}
where in the last line we have used the Dirac bracket \eqref{DB} together with the Dirac bracket \eqref{DB2} \cite{BodendorferCanonicalStructureOf}. Recalling now that $P^{kl}=P^{kk}\delta^{kl}$ and $q_{mn}=q_{mm}\delta_{mn}$, \eqref{cpDB} gives

\be
\left\{c^i(q,P),p_j(q,P)\right\}_*=-\frac{\partial c^i}{\partial P^{kk}}\left(\delta_m^k-q_{mm}q^{kk}\,\frac{f''(\chi)}{3f''(\chi)-2}\right)\frac{\partial p_j}{\partial q_{mm}}\;,
\ee
from which, using the expressions \eqref{eq:ps} for $p_j(q)$ and \eqref{fCM} for the function $f(\chi)$, it follows that
\be\label{cDB}
\begin{cases}
\left\{c^i(q,P),p_1(q)\right\}_*=-\frac{1}{2}\sqrt{\frac{q_{33}}{q_{22}}}\frac{\partial c^i}{\partial P^{22}}-\frac{1}{2}\sqrt{\frac{q_{22}}{q_{33}}}\frac{\partial c^i}{\partial P^{33}}+\frac{\sqrt{q_{22}q_{33}}}{3}\left(1-\sqrt{1-\frac{2}{3}\frac{\chi^2}{\chi_m^2}}\right)\sum_{k=1}^3\frac{1}{q_{kk}}\frac{\partial c^i}{\partial P^{kk}}\\
\left\{c^i(q,P),p_2(q)\right\}_*=-\frac{1}{2}\sqrt{\frac{q_{33}}{q_{11}}}\frac{\partial c^i}{\partial P^{11}}-\frac{1}{2}\sqrt{\frac{q_{11}}{q_{33}}}\frac{\partial c^i}{\partial P^{33}}+\frac{\sqrt{q_{11}q_{33}}}{3}\left(1-\sqrt{1-\frac{2}{3}\frac{\chi^2}{\chi_m^2}}\right)\sum_{k=1}^3\frac{1}{q_{kk}}\frac{\partial c^i}{\partial P^{kk}}\\
\left\{c^i(q,P),p_3(q)\right\}_*=-\frac{1}{2}\sqrt{\frac{q_{22}}{q_{11}}}\frac{\partial c^i}{\partial P^{11}}-\frac{1}{2}\sqrt{\frac{q_{11}}{q_{22}}}\frac{\partial c^i}{\partial P^{22}}+\frac{\sqrt{q_{11}q_{22}}}{3}\left(1-\sqrt{1-\frac{2}{3}\frac{\chi^2}{\chi_m^2}}\right)\sum_{k=1}^3\frac{1}{q_{kk}}\frac{\partial c^i}{\partial P^{kk}} 
\end{cases} 
\ee
Let us focus on the case $i=1$. Imposing the condition \eqref{PBDB1} for the Dirac brackets above, it yields the following system of first-order PDEs for the unknown function $c^1(q,P)$
\be\label{PDE1}
\begin{pmatrix}
A_1 & A_2 & A_3\\
B_1 & B_2 & B_3\\
C_1 & C_2 & C_3
\end{pmatrix}\begin{pmatrix}
\frac{\partial c^1}{\partial P^{11}}\\
\frac{\partial c^1}{\partial P^{22}}\\
\frac{\partial c^1}{\partial P^{33}}
\end{pmatrix}=\begin{pmatrix}
1\\
0\\
0
\end{pmatrix}\;,
\ee
where, according to \eqref{cDB}, the matrix of coefficients is given by
\be\label{coeffPDE1}
\begin{pmatrix}
\frac{1}{3}\frac{\sqrt{q_{22}q_{33}}}{q_{11}}\left(1-\sqrt{1-\frac{2}{3}\frac{\chi^2}{\chi_m^2}}\right) & -\frac{1}{3}\sqrt{\frac{q_{33}}{q_{22}}}\left(\frac{1}{2}+\sqrt{1-\frac{2}{3}\frac{\chi^2}{\chi_m^2}}\right) & -\frac{1}{3}\sqrt{\frac{q_{22}}{q_{33}}}\left(\frac{1}{2}+\sqrt{1-\frac{2}{3}\frac{\chi^2}{\chi_m^2}}\right)\\
-\frac{1}{3}\sqrt{\frac{q_{33}}{q_{11}}}\left(\frac{1}{2}+\sqrt{1-\frac{2}{3}\frac{\chi^2}{\chi_m^2}}\right) & \frac{1}{3}\frac{\sqrt{q_{11}q_{33}}}{q_{22}}\left(1-\sqrt{1-\frac{2}{3}\frac{\chi^2}{\chi_m^2}}\right) & -\frac{1}{3}\sqrt{\frac{q_{11}}{q_{33}}}\left(\frac{1}{2}+\sqrt{1-\frac{2}{3}\frac{\chi^2}{\chi_m^2}}\right)\\
-\frac{1}{3}\sqrt{\frac{q_{22}}{q_{11}}}\left(\frac{1}{2}+\sqrt{1-\frac{2}{3}\frac{\chi^2}{\chi_m^2}}\right) & -\frac{1}{3}\sqrt{\frac{q_{11}}{q_{22}}}\left(\frac{1}{2}+\sqrt{1-\frac{2}{3}\frac{\chi^2}{\chi_m^2}}\right) & \frac{1}{3}\frac{\sqrt{q_{11}q_{22}}}{q_{33}}\left(1-\sqrt{1-\frac{2}{3}\frac{\chi^2}{\chi_m^2}}\right)
\end{pmatrix}\;.
\ee
Solving \eqref{PDE1} for the derivatives $\frac{\partial c^1}{\partial P^{kk}}$, we get
\begin{align}
&\frac{\partial c^1}{\partial P^{11}}=\frac{1}{3}\frac{q_{11}}{\sqrt{q_{22}q_{33}}}\left(4-\frac{1}{\sqrt{1-\frac{2}{3}\frac{\chi^2}{\chi_m^2}}}\right)\;,\label{dc1dP11}\\
&\frac{\partial c^1}{\partial P^{22}}=-\frac{2}{3}\sqrt{\frac{q_{22}}{q_{33}}}\left(1+\frac{1}{2}\frac{1}{\sqrt{1-\frac{2}{3}\frac{\chi^2}{\chi_m^2}}}\right)\label{dc1dP22}\;,\\
&\frac{\partial c^1}{\partial P^{33}}=-\frac{2}{3}\sqrt{\frac{q_{33}}{q_{22}}}\left(1+\frac{1}{2}\frac{1}{\sqrt{1-\frac{2}{3}\frac{\chi^2}{\chi_m^2}}}\right)\;,\label{dc1dP33}
\end{align}
which, recalling the expression \eqref{chi} for $\chi$, can be integrated to
\be\label{c1}
c^1(q,P) = G_1(q) + \frac{4}{3} \frac{q_{11}}{\sqrt{q_{22} q_{33}}} P^{11} - \frac{2}{3} \sqrt{\frac{q_{33}}{q_{22}}} P^{33} - \frac{2}{3} \sqrt{\frac{q_{22}}{q_{33}}} P^{22} + \frac{\chi_m}{\sqrt{6}} \sqrt{q_{11}} \arcsin\left(\sqrt{\frac{2}{3}} \frac{\chi}{\chi_m}\right)\;. 
\ee
Similar expressions are obtained for $c^2$ and $c^3$:
\be\label{c2}
c^2(q,P) = G_2(q) + \frac{4}{3} \frac{q_{22}}{\sqrt{q_{11} q_{33}}} P^{22} - \frac{2}{3} \sqrt{\frac{q_{33}}{q_{11}}} P^{33} - \frac{2}{3} \sqrt{\frac{q_{11}}{q_{33}}} P^{11} + \frac{\chi_m}{\sqrt{6}} \sqrt{q_{22}} \arcsin\left( \sqrt{\frac{2}{3}} \frac{\chi}{\chi_m}\right)\;, 
\ee
\be\label{c3}
c^3(q,P) = G_3(q) + \frac{4}{3} \frac{q_{33}}{\sqrt{q_{11} q_{22}}} P^{33} - \frac{2}{3} \sqrt{\frac{q_{11}}{q_{22}}} P^{11} - \frac{2}{3} \sqrt{\frac{q_{22}}{q_{11}}} P^{22} + \frac{\chi_m}{\sqrt{6}} \sqrt{q_{33}} \arcsin\left( \sqrt{\frac{2}{3}} \frac{\chi}{\chi_m} \right)\;. 
\ee

It is possible to leave the functions $G_i(q)$ free and to determine a differential equation for them by demanding $\left\{c^i(q,P),c^j(q,P)\right\}_*=0$. A comparison to loop quantum cosmology on the other hand tells us to set $G_i=0$, as $c^i \rightarrow 0$ for $\dot p_i \rightarrow 0$ for generic polymerisation functions consistent with the classical limit\footnote{One needs to demand that $f(c) \sim c$ as $c \rightarrow 0$ for any good polymerisation function $f$, as otherwise the classical dynamics is changed already for low curvatures.}. We will thus drop $G_i$ in the following. 

As shown in the appendix, the condition \eqref{PBDB2} for the Dirac brackets between the $c^i$ given in \eqref{c1}-\eqref{c3} is satisfied and \eqref{c1}-\eqref{c3} with $G_i=0$ thus provide the desired solution for $c^i(q,P)$.

\subsection{Hamiltonian in LQC-variables}

The relations $p_i(q)$ and $c^i(q,P)$ found in the previous section now have to be inverted in order to insert the resulting expressions for the $q_{ii}$ and $P^{ii}$ in terms of the $c^i,p_i$ into the Hamiltonian \eqref{HCM3}. Inverting the functions $p_i(q)$ is straight forward:
\be\label{qofp}
q_{11}=\frac{p_2p_3}{p_1},\quad q_{22}=\frac{p_1p_3}{p_2},\quad q_{33}=\frac{p_1p_2}{p_3}\;.
\ee
On the other hand, inverting $c^i(q,P)$ is not trivial since also $\chi$ depends on $P^{ii}$. Therefore, let us first consider 
\be
\frac{\eqref{c1}}{\sqrt{q_{11}}} + \frac{\eqref{c2}}{\sqrt{q_{22}}} + \frac{\eqref{c3}}{\sqrt{q_{33}}} = \sum_{i=1 }^3 \frac{c^i}{\sqrt{q_{ii}}} = \sqrt{\frac{3}{2}} \chi_m \arcsin\left( \sqrt{\frac{2}{3}} \frac{\chi}{\chi_m} \right)\;,
\ee
from which, we find 
\be
\chi = \sqrt{\frac{3}{2}} \chi_m \sin\left(\sqrt{\frac{2}{3}} \frac{1}{\chi_m} \sum_{i=1}^3 \frac{c^i}{\sqrt{q_{ii}}} \right) \stackrel{\eqref{qofp}}{=} \sqrt{\frac{3}{2}} \chi_m \sin\left(\sqrt{\frac{2}{3}} \frac{1}{\chi_m} \frac{p_i c^i}{\sqrt{p_1p_2p_3}} \right) \;.
\ee
Inserting this back into \eqref{c1}-\eqref{c3} and using also \eqref{qofp}, we get a system of linear equations in the $P^{ii}$:
\begin{equation}\label{eqsys}
\begin{pmatrix}
4 \frac{p_2 p_3}{p_1^2} & -2 \frac{p_3}{p_2} & -2 \frac{p_2}{p_3} \\
-2 \frac{p_3}{p_1} & 4 \frac{p_1 p_3}{p_2^2} & -2 \frac{p_1}{p_3} \\
-2 \frac{p_2}{p_1} & -2 \frac{p_1}{p_2} & 4 \frac{p_1 p_2}{p_3^2}
\end{pmatrix}
\begin{pmatrix}
P^{11} \\
P^{22} \\
P^{33}
\end{pmatrix}
=
\begin{pmatrix}
2 c^1 - \frac{p_2}{p_1} c^2 - \frac{p_3}{p_1} c^3 \\
2 c^2 - \frac{p_3}{p_2} c^3 - \frac{p_1}{p_2} c^1 \\
2 c^3 - \frac{p_2}{p_3} c^2 - \frac{p_1}{p_3} c^1 
\end{pmatrix}\;.
\end{equation}
This system of equations is underdetermined since the determinant of the left-most matrix vanishes. Indeed, solving the second equation of \eqref{eqsys} for $P^{22}$, we have
\begin{equation}\label{P2}
P^{22} = \frac{1}{4} \frac{p_2^2}{p_1 p_3} \left( 2c^2 - \frac{p_3}{p_2} c^3 - \frac{p_1}{p_2} c^1 + 2 \frac{p_3}{p_1} P^{11} + 2 \frac{p_1}{p_3} P^{33} \right)\;,
\end{equation}
and, inserting it into the third equation of \eqref{eqsys}, it yields
\begin{equation}\label{P33}
P^{33} = - \frac{p_3}{2p_2}c^1 + \frac{p_3^2}{2 p_1 p_2}c^3 + \frac{p_3^2}{p_1^2} P^{11}\; .
\end{equation}
Inserting this back into \eqref{P2}, we get
\begin{equation}\label{P22}
P^{22} = - \frac{p_2}{2p_3}c^1 + \frac{p_2^2}{2 p_1 p_3}c^2 + \frac{p_2^2}{p_1^2} P^{11}\; .
\end{equation}
Inserting both back into the first equation of \eqref{eqsys} solves it identically, thus showing that the system is underdetermined. Nevertheless, the condition
\begin{align}
&-\sqrt{\frac{3}{2}} \chi_m \sqrt{p_1 p_2 p_3} \sin\left(\sqrt{\frac{2}{3}} \frac{1}{\chi_m} \frac{p_i c^i}{\sqrt{p_1p_2p_3}} \right)  \nonumber \\ &= -\sqrt{q} \chi \overset{\eqref{chi}}{=} q_{ii} P^{ii} = 3 \frac{p_2 p_3}{p_1} P^{11} - p_1 c^1 + \frac{1}{2} p_2 c^2 + \frac{1}{2} p_3 c^3
\end{align}
fixes $P^{11}$ and hence $P^{22}$ and $P^{33}$ given in \eqref{P33}, \eqref{P22} to be
\begin{align}
P^{11} =& - \frac{\chi_m}{\sqrt{6}} \frac{p_1}{p_2 p_3} \sqrt{p_1 p_2 p_3} \sin\left( \sqrt{\frac{2}{3}} \frac{1}{\chi_m} \frac{p_i c^i}{\sqrt{p_1p_2p_3}} \right) + \frac{1}{3} \frac{p_1^2}{p_2 p_3} c^1 - \frac{1}{6} \frac{p_1}{p_3} c^2 - \frac{1}{6} \frac{p_1}{p_2} c^3\;, \\
P^{22} =& - \frac{\chi_m}{\sqrt{6}} \frac{p_2}{p_1 p_3} \sqrt{p_1 p_2 p_3} \sin\left( \sqrt{\frac{2}{3}} \frac{1}{\chi_m} \frac{p_i c^i}{\sqrt{p_1p_2p_3}} \right) + \frac{1}{3} \frac{p_2^2}{p_1 p_3} c^2 - \frac{1}{6} \frac{p_2}{p_3} c^1 - \frac{1}{6} \frac{p_2}{p_1} c^3\;, \\
P^{33} =& - \frac{\chi_m}{\sqrt{6}} \frac{p_3}{p_1 p_2} \sqrt{p_1 p_2 p_3} \sin\left( \sqrt{\frac{2}{3}} \frac{1}{\chi_m} \frac{p_i c^i}{\sqrt{p_1p_2p_3}} \right) + \frac{1}{3} \frac{p_3^2}{p_1 p_2} c^3 - \frac{1}{6} \frac{p_3}{p_2} c^1 - \frac{1}{6} \frac{p_3}{p_1} c^2\;.
\end{align}
These expressions can now be inserted into the Hamiltonian \eqref{HCM3}, giving 
\begin{align}
\mathcal{H}_{grav} =&-\frac{1}{\sqrt{p_1p_2p_3}} \left( p_1c^1 p_2c^2 + p_1c^1 c^3p_3 + p_2c^2p_3c^3\right) 
\notag
\\&-2\sqrt{p_1p_2p_3} \chi_m^2 \sin^2\left(\frac{1}{\sqrt{6} \chi_m} \frac{p_i c^i}{\sqrt{p_1p_2p_3}} \right)  +\frac{1}{3}\sqrt{p_1p_2p_3}\left(\frac{p_i c^i}{\sqrt{p_1p_2p_3}}\right)^2\;, \label{HCMcp}
\end{align}
or equivalently
\begin{align} \label{eq:FinalH}
\mathcal{H}_{grav}=&-\sqrt{p_1 p_2 p_3} \left(\frac{ c^2 c^3}{p_1} +\frac{ c^3 c^1}{p_2} +\frac{ c^1 c^2}{p_3} \right) - 2 \sqrt{p_1p_2p_3} \chi_m^2 \left( \sin^2(x) - x^2 \right) , \; x = \frac{1}{\sqrt{6} \chi_m} \frac{p_i c^i}{\sqrt{p_1p_2p_3}}.
\end{align}
Taking the last expression, it is easy to check that the classical theory is obtained in the limit $\chi_m \rightarrow \infty$, hence $x\rightarrow 0$, where the $c^i$ also assume their classical meaning.

\eqref{eq:FinalH} puts us into a position to answer the central question of the paper. While the $\sin^2(x)$ term can be written as a linear combination of point holonomies and therefore arise as a part of an effective loop quantum cosmology Hamiltonian, the other terms, being linear or quadratic in $c^i$, cannot. We thus conclude that already in the context of Bianchi I models, limiting curvature mimetic gravity cannot be written as an effective loop quantum cosmology type model. 

It also becomes clear at this point that changing the function $f$ given in \eqref{fCM} cannot change this conclusion: since $f$ depends only on $\chi$ and thus a certain combination of the $c^i$, we cannot turn all individual terms containing $c^i$ in \eqref{eq:BianchiIH} into almost-periodic functions by a modification of the Hamiltonian as in \eqref{HCM3}. 

To reconcile this conclusion with the previous work, let us consider the isotropic limit in which $p_1 = p_2 = p_3=p$ and $c^1=c^2=c^3=c$. The Hamiltonian \eqref{HCMcp} then reduces to
\be
\mathcal{H}_{grav} = -2\sqrt{p^3} \chi_m^2 \sin^2\left(\sqrt{\frac{3}{2}}\frac{1}{\chi_m} \frac{c}{\sqrt{p}}\right).
\ee
For the isotropic case, the relations 
\be
c = -\frac{\sqrt{p}}{3} b \, , \qquad v = \sqrt{p^3}
\ee
hold. Thus, the Hamiltonian can be written as 
\begin{equation}
\mathcal{H}_{grav} = -2 v \chi_m^2 \sin^2\left(\frac{1}{\sqrt{6} \chi_m} b\right) = -\frac{1}{3} v \frac{\sin^2\left(\lambda b\right)}{\lambda^2} \label{eq:IsotropicH}
\end{equation}
with polymerisation scale $\lambda = \frac{1}{\sqrt{6} \chi_m}$, in agreement with \cite{BodendorferCanonicalStructureOf, NouiEffectiveLoopQuantum}. The problematic terms in \eqref{eq:FinalH} drop and the above argument fails since \eqref{eq:IsotropicH} depends only on one gravitational variable.

\section{Conclusion} \label{sec:Conclusion}

In this paper, we have investigated whether limiting curvature mimetic gravity can be interpreted as effective loop quantum cosmology beyond the homogeneous and isotropic setting. Our result was negative, owing to the appearance of terms in the Hamiltonian constraint that cannot be implemented as well-defined operators on the loop quantum cosmology Hilbert space. While the computation was based on a specific choice of potential leading to the limiting curvature feature, our conclusion should hold for general potentials as explained above. Despite the differences encountered between effective loop quantum cosmology and limiting curvature mimetic gravity, it is still interesting to consider such models and generalisations thereof, see e.g. \cite{NouiEffectiveLoopQuantum}, as toy models for loop quantum gravity. This is due to a similar qualitative behaviour in the homogeneous, but non-isotropic context such as the presence of the bounce interpolating between two classical Bianchi I spacetimes as well as the same rules relating the Kasner exponents before and after the bounce, cfr. Eq. (45) in \cite{ChamseddineResolvingCosmologicalSingularities} and Eq. (40) in \cite{WilsonEwingTheLoopQuantum} (see however \cite{BrahmaOnSingularityResolution} for some subtleties). It would be interesting to continue this comparison by dropping the homogeneity assumption and proceed to compare black hole solutions \cite{ChamseddineNonsingularBlackHole, CorichiLoopQuantizationOf} as initiated in \cite{AchourNonSingularBlack}.

As a byproduct, the variable transformation derived in this paper brought the Dirac bracket for Bianchi I models in limiting curvature mimetic gravity into canonical form. Independently of a relation to a polymer quantisation, we may now take this choice of variable as a starting point for a treatment \`a la Wheeler-de Witt.  
  
\section*{Acknowledgments}

The authors were supported by an International Junior Research Group grant of the Elite Network of Bavaria. Discussions with Andreas Sch\"afer and John Schliemann are gratefully acknowledged.

\begin{appendix}

\section{Dirac bracket $\left\{c^i(q,P),c^j(q,P)\right\}_*$}

We have to impose the condition \eqref{PBDB2} for the Dirac brackets between the $c^i$ given in \eqref{c1}-\eqref{c3}. Therefore, by using the Dirac brackets \eqref{DB}, \eqref{DB1} and the expression \eqref{fCM} for $f(\chi)$, the l.h.s. of \eqref{PBDB2} gives
\be\label{cDB2}
\begin{split}
\{c^i&(q,P),c^j(q,P)\}_*=\frac{\partial c^i}{\partial q_{kl}}\{q_{kl},P^{mn}\}_*\frac{\partial c^j}{\partial P^{mn}}+\frac{\partial c^i}{\partial P^{mn}}\{P^{mn},q_{kl}\}_*\frac{\partial c^j}{\partial q_{kl}}+\frac{\partial c^i}{\partial P^{kl}}\{P^{kl},P^{mn}\}_*\frac{\partial c^j}{\partial P^{mn}}\\
&=\left(\frac{\partial c^i}{\partial q_{kl}}\frac{\partial c^j}{\partial P^{mn}}-\frac{\partial c^j}{\partial q_{kl}}\frac{\partial c^i}{\partial P^{mn}}\right)\{q_{kl},P^{mn}\}_*+\frac{\partial c^i}{\partial P^{kl}}\frac{\partial c^j}{\partial P^{mn}}\{P^{kl},P^{mn}\}_*\\
&=\left(\frac{\partial c^i}{\partial q_{kl}}\frac{\partial c^j}{\partial P^{mn}}-\frac{\partial c^j}{\partial q_{kl}}\frac{\partial c^i}{\partial P^{mn}}\right)\left(\delta_{(k}^m\delta_{l)}^n-\frac{1}{3}q_{kl}q^{mn}\left(1-\sqrt{1-\frac{2}{3}\frac{\chi^2}{\chi_m^2}}\right)\right)\\
&\quad-\frac{\partial c^i}{\partial P^{kl}}\frac{\partial c^j}{\partial P^{mn}}\left(q^{kl}P^{mn}-q^{mn}P^{kl}\right)\frac{1}{3}\left(1-\sqrt{1-\frac{2}{3}\frac{\chi^2}{\chi_m^2}}\right)\\
&=\left(\frac{\partial c^i}{\partial q_{kk}}\frac{\partial c^j}{\partial P^{kk}}-\frac{\partial c^j}{\partial q_{kk}}\frac{\partial c^i}{\partial P^{kk}}\right)-\frac{1}{3}\left(1-\sqrt{1-\frac{2}{3}\frac{\chi^2}{\chi_m^2}}\right)\\
&\quad\cdot\left(\left(\frac{\partial c^i}{\partial q_{kk}}\frac{\partial c^j}{\partial P^{mm}}-\frac{\partial c^j}{\partial q_{kk}}\frac{\partial c^i}{\partial P^{mm}}\right)q_{kk}q^{mm}+\left(\frac{\partial c^i}{\partial P^{kk}}\frac{\partial c^j}{\partial P^{mm}}-\frac{\partial c^j}{\partial P^{kk}}\frac{\partial c^i}{\partial P^{mm}}\right)q^{kk}P^{mm}\right)\;.
\end{split}
\ee
Let us compute the three sums involving the derivatives of the $c^i$ separately. For instance for $i=1$ and $j=2$, using the expressions \eqref{c1}, \eqref{c2} for $c^1$, $c^2$ and \eqref{chi} for $\chi$, we have

\begin{align}
&\frac{\partial c^1}{\partial q_{11}}=\frac{4}{3}\frac{P^{11}}{\sqrt{q_{22}q_{33}}}+\frac{\chi_m}{2\sqrt{6}}\frac{1}{\sqrt{q_{11}}}\arcsin\left( \sqrt{\frac{2}{3}} \frac{\chi}{\chi_m}\right)\notag\\
&\qquad\quad+\frac{\sqrt{q_{11}}}{3}\frac{1}{\sqrt{1-\frac{2}{3}\frac{\chi^2}{\chi_m^2}}}\left(-\frac{1}{2}\frac{P^{11}}{\sqrt{q_{11}q_{22}q_{33}}}+\frac{1}{2q_{11}}\sqrt{\frac{q_{22}}{q_{11}q_{33}}}P^{22}+\frac{1}{2q_{11}}\sqrt{\frac{q_{33}}{q_{11}q_{22}}}P^{33}\right)\;,\label{dc1dq11}\\
&\frac{\partial c^1}{\partial q_{22}}=-\frac{2}{3}\frac{q_{11}P^{11}}{q_{22}\sqrt{q_{22}q_{33}}}+\frac{1}{3}\frac{1}{q_{22}}\sqrt{\frac{q_{33}}{q_{22}}}P^{33}-\frac{1}{3}\frac{P^{22}}{\sqrt{q_{22}q_{33}}}\notag\\
&\qquad\quad+\frac{\sqrt{q_{11}}}{3}\frac{1}{\sqrt{1-\frac{2}{3}\frac{\chi^2}{\chi_m^2}}}\left(-\frac{1}{2}\frac{P^{22}}{\sqrt{q_{11}q_{22}q_{33}}}+\frac{1}{2q_{22}}\sqrt{\frac{q_{11}}{q_{22}q_{33}}}P^{11}+\frac{1}{2q_{22}}\sqrt{\frac{q_{33}}{q_{11}q_{22}}}P^{33}\right)\;,\label{dc1dq22}\\
&\frac{\partial c^1}{\partial q_{33}}=-\frac{2}{3}\frac{q_{11}P^{11}}{q_{33}\sqrt{q_{22}q_{33}}}-\frac{1}{3}\frac{P^{33}}{\sqrt{q_{22}q_{33}}}+\frac{1}{3}\frac{1}{q_{33}}\sqrt{\frac{q_{22}}{q_{33}}}P^{22}\notag\\
&\qquad\quad+\frac{\sqrt{q_{11}}}{3}\frac{1}{\sqrt{1-\frac{2}{3}\frac{\chi^2}{\chi_m^2}}}\left(-\frac{1}{2}\frac{P^{33}}{\sqrt{q_{11}q_{22}q_{33}}}+\frac{1}{2q_{33}}\sqrt{\frac{q_{11}}{q_{22}q_{33}}}P^{11}+\frac{1}{2q_{33}}\sqrt{\frac{q_{22}}{q_{11}q_{33}}}P^{22}\right)\label{dc1dq33}\;,\\
&\frac{\partial c^2}{\partial q_{11}}=-\frac{2}{3}\frac{q_{22}P^{22}}{q_{11}\sqrt{q_{11}q_{33}}}+\frac{1}{3}\frac{1}{q_{11}}\sqrt{\frac{q_{33}}{q_{11}}}P^{33}-\frac{1}{3}\frac{P^{11}}{\sqrt{q_{11}q_{33}}}\notag\\
&\qquad\quad+\frac{\sqrt{q_{22}}}{3}\frac{1}{\sqrt{1-\frac{2}{3}\frac{\chi^2}{\chi_m^2}}}\left(-\frac{1}{2}\frac{P^{11}}{\sqrt{q_{11}q_{22}q_{33}}}+\frac{1}{2q_{11}}\sqrt{\frac{q_{22}}{q_{11}q_{33}}}P^{22}+\frac{1}{2q_{11}}\sqrt{\frac{q_{33}}{q_{11}q_{22}}}P^{33}\right)\label{dc2dq11}\;,\\
&\frac{\partial c^2}{\partial q_{22}}=\frac{4}{3}\frac{P^{22}}{\sqrt{q_{11}q_{33}}}+\frac{\chi_m}{2\sqrt{6}}\frac{1}{\sqrt{q_{22}}}\arcsin\left( \sqrt{\frac{2}{3}} \frac{\chi}{\chi_m}\right)\notag\\
&\qquad\quad+\frac{\sqrt{q_{22}}}{3}\frac{1}{\sqrt{1-\frac{2}{3}\frac{\chi^2}{\chi_m^2}}}\left(-\frac{1}{2}\frac{P^{22}}{\sqrt{q_{11}q_{22}q_{33}}}+\frac{1}{2q_{22}}\sqrt{\frac{q_{11}}{q_{22}q_{33}}}P^{11}+\frac{1}{2q_{22}}\sqrt{\frac{q_{33}}{q_{11}q_{22}}}P^{33}\right)\;,\label{dc2dq22}
\end{align}
\begin{align}
&\frac{\partial c^2}{\partial q_{33}}=-\frac{2}{3}\frac{q_{22}P^{22}}{q_{33}\sqrt{q_{11}q_{33}}}-\frac{1}{3}\frac{P^{33}}{\sqrt{q_{11}q_{33}}}+\frac{1}{3}\frac{1}{q_{33}}\sqrt{\frac{q_{11}}{q_{33}}}P^{11}\notag\\
&\qquad\quad+\frac{\sqrt{q_{22}}}{3}\frac{1}{\sqrt{1-\frac{2}{3}\frac{\chi^2}{\chi_m^2}}}\left(-\frac{1}{2}\frac{P^{33}}{\sqrt{q_{11}q_{22}q_{33}}}+\frac{1}{2q_{33}}\sqrt{\frac{q_{22}}{q_{11}q_{33}}}P^{22}+\frac{1}{2q_{33}}\sqrt{\frac{q_{11}}{q_{22}q_{33}}}P^{11}\right)\;,\label{dc2dq33}
\end{align}
\begin{align}
&\frac{\partial c^2}{\partial P^{11}}=-\frac{2}{3}\sqrt{\frac{q_{11}}{q_{33}}}\left(1+\frac{1}{2}\frac{1}{\sqrt{1-\frac{2}{3}\frac{\chi^2}{\chi_m^2}}}\right)\;,\label{dc2dP11}\\
&\frac{\partial c^2}{\partial P^{22}}=\frac{1}{3}\frac{q_{22}}{\sqrt{q_{11}q_{33}}}\left(4-\frac{1}{\sqrt{1-\frac{2}{3}\frac{\chi^2}{\chi_m^2}}}\right)\;,\label{dc2dP22}\\
&\frac{\partial c^2}{\partial P^{33}}=-\frac{2}{3}\sqrt{\frac{q_{33}}{q_{11}}}\left(1+\frac{1}{2}\frac{1}{\sqrt{1-\frac{2}{3}\frac{\chi^2}{\chi_m^2}}}\right)\;,\label{dc2dP33}
\end{align}

\noindent
from which, together with \eqref{dc1dP11}-\eqref{dc1dP33}, it follows that
\begin{align}
&\frac{\partial c^1}{\partial q_{kk}}\frac{\partial c^2}{\partial P^{kk}}-\frac{\partial c^2}{\partial q_{kk}}\frac{\partial c^1}{\partial P^{kk}}=-\frac{2}{3}\frac{1}{q_{33}}\left(1-\frac{1}{\sqrt{1-\frac{2}{3}\frac{\chi^2}{\chi_m^2}}}\right)\left(\sqrt{\frac{q_{11}}{q_{22}}}P^{11}-\sqrt{\frac{q_{22}}{q_{11}}}P^{22}\right),\label{first}\\
&\left(\frac{\partial c^1}{\partial q_{kk}}\frac{\partial c^2}{\partial P^{mm}}-\frac{\partial c^2}{\partial q_{kk}}\frac{\partial c^1}{\partial P^{mm}}\right)q_{kk}q^{mm}=0,\label{second}\\
&\left(\frac{\partial c^1}{\partial P^{kk}}\frac{\partial c^2}{\partial P^{mm}}-\frac{\partial c^2}{\partial P^{kk}}\frac{\partial c^1}{\partial P^{mm}}\right)q^{kk}P^{mm}=\frac{1}{q_{33}}\frac{2}{\sqrt{1-\frac{2}{3}\frac{\chi^2}{\chi_m^2}}}\left(\sqrt{\frac{q_{11}}{q_{22}}}P^{11}-\sqrt{\frac{q_{22}}{q_{11}}}P^{22}\right)\;.\label{third}
\end{align}
Using \eqref{first}-\eqref{third} and similar expressions for the other Dirac brackets, we find that \eqref{cDB2} is satisfied and that \eqref{c1}, \eqref{c2}, and \eqref{c3} provide the desired solution for $c^i(q,P)$.

\end{appendix}


\end{document}